\documentclass[preprint,12pt]{aastex}
\begin{document}

\title{The Correlation between X-Ray Line Ionization and Optical Spectral
Types of the OB Stars}

\author{Nolan R. Walborn}
\affil{Space Telescope Science Institute\altaffilmark{1}, 3700 San Martin 
Drive, Baltimore, MD 21218, USA}
\authoremail{walborn@stsci.edu}

\author{Joy S. Nichols}
\affil{Harvard-Smithsonian Center for Astrophysics, 60 Garden Street,
Cambridge, MA 02138}
\authoremail{jnichols@cfa.harvard.edu}

\author{Wayne L. Waldron}
\affil{Eureka Scientific, Inc., 2452 Delmer Street, Suite 100, Oakland, 
CA 94602}
\authoremail{wwaldron@satx.rr.com}

\altaffiltext{1}{Operated by the Association of Universities for Research
in Astronomy, Inc., under NASA contract NAS 5-26555.}

\begin{abstract}
Marked correlations are reported between the ionization of the X-ray line 
spectra of normal OB stars, as observed by the \textit{Chandra X-Ray 
Observatory}, and their optical spectral types.  These correlations include 
the progressive weakening of the higher ionization relative to the lower 
ionization X-ray lines with advancing spectral type, and the similarly 
decreasing intensity ratios of the H-like to He-like lines of the $\alpha$ 
ions.  These relationships were not predicted by models, nor have they been clearly evident in astrophysical studies of a few objects; rather, they have emerged from morphological analysis of an adequate (albeit still small) sample, from which known peculiar objects such as magnetic stars and very rapid rotators have been isolated to reveal the normal trends.  This process is analogous to that which first demonstrated the strong relationships between the UV wind profiles and the optical spectral types of normal OB stars, which likely bear a physical as well as a historical connection to the present X-ray results.  Since the optical spectral types are calibrated in terms of fundamental stellar parameters, it follows that the winds and X-ray spectra are determined by the latter.  These observations provide strong guidance for further astrophysical modeling of these phenomena.
\end{abstract}


\section{Introduction}

When a new observational domain, comprising a diverse phenomenology not yet
fully explained or predicted by physical models, is opened to investigation, 
the science of morphology may provide a useful initial formulation.  For
instance, morphology can discern empirical trends, separate the normal  
(majority) behavior from peculiar exceptions, reveal correlations with
other kinds of information, and suggest or eliminate hypotheses for 
subsequent physical analysis, all with a minimum of prior assumptions.  
A frequent occurrence of this situation in astronomy is access to a new 
wavelength region or data of higher information content.  A specific example from the relatively recent past is the extensive database of high-resolution ultraviolet stellar spectra generated by the \textit{International Ultraviolet Explorer} (\textit{IUE}) satellite.  During the early 1980s, there was a debate about the relationship if any of OB stellar winds to the fundamental stellar parameters.  A majority of specialists believed that it was weak or nonexistent, because of the large scatter in plots of
mass-loss rate versus effective temperature or luminosity; however, it is 
now known that the derivations of those quantities were inadequate at that 
time.  The debate was settled by a purely morphological demonstration of 
the detailed correlations between the wind profiles in the \textit{IUE} data 
and the optical spectral types, which had in turn been calibrated in terms of fundamental parameters (Walborn et~al.\ 1985, 1995, 2009; Rountree \& 
Sonneborn 1991, 1993; Penny et~al.\ 1996).  The database provided by the 
\textit{Far Ultraviolet Spectroscopic Explorer} (\textit{FUSE}) supported 
the extension of those correlations to the rich spectral region below 
Lyman-$\alpha$ (Walborn et~al.\ 2002, Pellerin et~al.\ 2002).  Of course, 
morphology does not explain anything, but it may pave the way toward the 
ultimate objective of physical understanding.

The \textit{Chandra} (and \textit{XMM-Newton}) X-ray satellites have provided the first stellar spectroscopic data of adequate quality to investigate the systematics of that regime.  The samples remain small to date because the required high-resolution data are time-consuming to acquire with those systems. Nevertheless, they are sufficient for an initial morphological survey, which is presented here based on the \textit{Chandra} data, with results that may be surprising.

\section{Observations}

All of the observations were obtained with the \textit{Chandra} Advanced CCD Imaging Spectrometer (ACIS) in its High Energy Transmission Grating 
Spectrometer (HETGS) configuration, which in turn simultaneously provides
Medium Energy Grating (MEG) and High Energy Grating (HEG) data.  The resolving powers are 660 at 15~\AA\ and 1000 at 12.4~\AA.  We have combined the two datasets whenever possible, i.e., whenever the HEG S/N was adequate.  The data discussed here were obtained from the TGCat (Huenemoerder et~al.\ 2009) web-based catalog of \textit{Chandra} grating data.  TGCat processing starts with the Level~1 event data downloaded from the \textit{Chandra} archive.  Updated calibrations and responses are applied to the data and the zeroth order position is determined to high accuracy.  The resulting TGCat Level~2 extracted spectral data were used to create the plots shown here.

Most of the data discussed here were archival, but three of the four 
targets observed in our \textit{Chandra} program (PI Waldron, ID 6200204) to fill gaps in the extant coverage of the HR Diagram are included.  They are HD~93250, 9~Sgr, and 15~Mon; our fourth target, HD~93129AB, presents a complex, composite spectrum that will be discussed separately (Nichols et~al., in preparation.)  Stellar and observational parameters are listed in Table~1, in the order in which the objects appear in Figures~1--6, together with extensive references to previous analyses of these stars.  Further explanation of several columns in the table is given below.

Other relevant data exist in the \textit{XMM-Newton} archive, which we
plan to explore.  Those observations provide better coverage of the CNO
lines at the longer wavelengths, and they include several stars not observed
by \textit{Chandra}.

\section{Results}

The 5--25~\AA\ X-ray spectra of 14 normal OB stars (or binary systems) are shown in order of advancing optical spectral types in Figures~1 and 2, which contain primarily main-sequence stars and supergiants, respectively, albeit with a few giants included to augment the coverage in each case.  In these plots, each X-ray spectrogram has been scaled to 1.0 at the peak emission feature.  Thus, absolute intensities cannot be directly compared between different objects, although line ratios can be.  Because of the large range of line intensities in the calibrated data, a complementary presentation of the same data, but with three different scalings at wavelengths below 11~\AA, from 11 to 18~\AA, and above 18~A is given in Figures~3 and~4. The three normalization factors for each object are listed in Table~1; their units are photons~s$^{-1}$~cm$^{-2}$~{\AA}$^{-1}$.  The largest of the three values (usually the third, except for three of the earliest types where it is the second) applies to Figures~1 and 2.  As indicated by the spectral types, several objects are known close binaries.  Parameters of the strong spectral lines are listed in Table~2.

Several correlations between the X-ray line ionization and the optical
spectral types are immediately apparent in these figures.  First, in
Figures~1 and 2, the higher ionization, shorter wavelength lines weaken
markedly relative to the lower ionization, longer wavelength lines in a
given spectrum along the sequences, although the oxygen lines are usually the strongest.  Second, in Figures~3 and 4, the behaviors of individual lines and line ratios, particularly those of the close pairs of lines from the H-like and He-like $\alpha$ ions, display strong trends along the sequences.  Specifically, S\,\textsc{xv} essentially disappears after O3.5 on the main sequence, and after O4 in the supergiants.  The Si\,\textsc{xiv}/Si\,\textsc{xiii} and Mg\,\textsc{xii}/Mg\,\textsc{xi} ratios diminish drastically 
between O3.5 and O4 in both sequences.  The Ne\,\textsc{x}/Ne\,\textsc{ix} ratio is the best diagnostic, because it is the most sensitive throughout this spectral-type range. Unfortunately, the Ne\,\textsc{x} feature is blended with an Fe\,\textsc{xvii} line that strengthens toward the later types (as do the other lines from the latter ion), interfering with the visual trend; this blend is elucidated at a larger scale in Figure~5.  Finally, the 
O\,\textsc{viii}/O\,\textsc{vii} trend is relatively weak throughout this range, although the ratio reverses between the extreme cases of HD~150136 and $\beta$~Cru (the spectrogram of HD~93250 has lower S/N and is unreliable at the longer wavelengths).  The large N/O ratios in $\zeta$~Pup and $\xi$~Per are most likely abundance effects in processed material, as previously derived in the former spectrum by Kahn et~al.\ (2001) in agreement with prior photospheric/wind analyses; see also Oskinova et~al.\ (2006).  Note that the \textit{differential} behaviors of the ionization ratios are themselves an ionization effect, with the intermediate Ne lines providing the best response to the relevant parameter range.  Quantitative measurements (and corresponding uncertainties) of these line ratios, as well as the implied temperatures, are given by Waldron \& Cassinelli (2007).  We present measurements of the Ne ratios below.

It is noteworthy that the binary nature of several objects, which
may have colliding winds, does not interfere with these observed trends,
although it may contribute to some of the scatter in them. Neither does the light to moderate range of interstellar extinctions among them (Table~1), for which no correction has been made here. The seven stars with E(B-V) between 0.3 and 0.5 are expected to have emission features longward of 18~\AA\ extincted by factors up to 2--3. One might be concerned that the earlier type normal stars tend to have the higher interstellar extinctions, but $\zeta$~Pup provides the counterexample with an early type and low extinction that fits the X-ray line sequence well.  Of course, wind or interstellar extinction always affects the longer wavelengths more, so it cannot cause the relatively weaker high ionization features at the later types.  Neither would it affect the ratios of the close pairs of H-like to He-like lines significantly.

The present sample is not adequate to investigate luminosity effects in detail, but there are a few suggestive indications of an ionization correlation with that dimension as well.  As noted above, the S\,\textsc{xv} line persists to a slightly later type in the supergiants.  More systematically, it can be seen that the Si\,\textsc{xiv} and Mg\,\textsc{xii} lines remain stronger at later types in the supergiants than on the main sequence (see also Waldron \& Cassinelli 2007).  Indeed, $\iota$~Ori rather breaks these trends in Figure~3, which may well be due to its giant nature.  More extensive coverage of intermediate luminosity classes with X-ray data of this quality is required to determine whether these effects correlate in detail with that dimension. 

Figure~6 displays the X-ray spectra of two magnetic and two very rapidly 
rotating OB stars.  It is easily seen that they have much higher ionizations 
than those of normal stars with the same spectral types.  They undoubtedly 
correspond to energetic circumstellar activity related to the fields or 
rotation (see also Wojdowski \& Schulz 2005, Zhekov \& Palla 2007, 
Westbrook et~al.\ 2008).\footnote{As shown in Table~1, $\zeta$~Oph and
$\gamma$~Cas have substantially higher {\it vsini} than the other stars
in this sample.  However, they are likely not typical rapid rotators, and
it should not be inferred that all objects in that category would have
anomalous X-ray spectra.  $\zeta$~Oph displays transient Oe/H$\alpha$ emission events (Niemel\"a \& M\'endez 1974, Ebbets 1981), while $\gamma$~Cas is subject to extreme shell episodes (Cowley \& Marlborough 1968) and may also be magnetic (Smith et al.\ 2004).}  This result emphasizes once again that it is essential to eliminate pathological cases from the samples, if one intends to investigate the normal phenomena.

To summarize and somewhat quantify our results, we have measured the optimal Ne ratio, by means of Gaussian fits to the lines.  Since the Ne\,\textsc{x} blend with the Fe\,\textsc{xvii} 12.124~\AA\ line cannot be resolved, we have corrected for it approximately by measuring the isolated Fe\,\textsc{xvii} 15.014~\AA\ line and subtracting it, scaled by the theoretical intensity ratio of 0.13, from the blend.  The resultant Ne ratios are listed in Table~1 and plotted versus spectral type in Figure~7, with the luminosity classes and the peculiar objects distinguished by different symbols; the trends described above are well shown.  We have also derived \mbox{X-ray} temperatures from the Ne ratios by procedures similar to those of Waldron \& Cassinelli (2007);\footnote{It should be noted that the present Ne\,\textsc{x}/Ne\,\textsc{ix} ratios are systematically smaller than those of Waldron \& Cassinelli (2007), for two main reasons: they chose not to remove the Fe\,\textsc{xvii} line from the Ne\,\textsc{x} blend, as was done here; and they estimated the Ne\,\textsc{ix} r-component in isolation from the f, i and other blended species, whereas we measured the entire complex.  However, the respective theoretical determinations approximate the different contents of the measurements, so that the temperatures derived from the two studies agree within $\pm 16\%$ in all cases except that of HD~206267 (35\%).
In this spectrum, the Fe\,\textsc{xvii} line at 15.014~\AA, which we use
to estimate the strength of the Fe\,\textsc{xvii} 12.124~\AA line blended
with Ne\,\textsc{x}, appears to be anomalously weak, leading to a
possible undercorrection and hence Ne\,\textsc{x} overestimate here 
(Waldron, in preparation).} they are likewise listed in Table~1 and 
plotted in Figure~7 against the stellar effective temperatures from the 
recalibration of the spectral types by Martins, Schaerer, \& Hillier (2005).  
The two plots are very similar, as expected from the good correlations 
between the respective variables.  No systematic dependence of the Ne X-ray ionization on the luminosity classes is evident in this sample, nor 
does a plot against the gravities show any meaningful trend (e.g., the normal class V stars span the full X-ray ionization range but have constant gravity).

\section{Discussion}

The systematic characteristics of OB X-ray spectra just described have not 
emerged clearly from previous studies.  The likely reasons are inadequate 
samples contaminated by peculiar objects, and the uncertainties of current
physical models for the phenomena.  As recounted in the Introduction, the
situation regarding the systematics of OB stellar winds during the early
1980s was remarkably similar, and it was substantially illuminated by
purely morphological investigations.  In fact, the mass-loss rates and
physical details of OB winds remain uncertain today, with the roles of
clumping, rotation, and possibly weak magnetic fields being actively
investigated.  Nevertheless, the purely empirical descriptions of the
phenomena stand as guides and objectives for the modeling.

The facts that the UV wind and X-ray spectral morphologies both correlate
with the optical spectral types is likely no coincidence.  Indeed, the X-rays most likely originate in the winds (Lucy \& White 1980, Owocki et~al.\ 1988, Feldmeier et al.\ 1997), and the observed correlations indicate that the detailed wind structures must be well determined by the fundamental stellar parameters, likely via the intense radiation fields of the OB stars.  Waldron \& Cassinelli (2007) measured and analyzed several parameters in the X-ray spectra of nearly the same sample discussed here.  The physical details are very complex, e.g., the lines considered here must form at different depths in the winds and are thus subject to differing amounts of internal extinction.  X-ray emission from shocks related to clumps distributed throughout the winds is discussed by Feldmeier et~al.\ 2003, Oskinova et~al.\ (2004, 2006), Zhekov \& Palla (2007), Cassinelli et~al.\ (2008), and Waldron \& Cassinelli (2009).  There may also be global asymmetric wind structures even in the normal stars, e.g., two-component equatorial vs.\ polar winds related to rotation and/or weak magnetic fields (Mullan \& Waldron 2006).  Nevertheless, the global systematics evident in the emergent spectra strongly suggest that all of these physical effects must be correlated and determined by the stellar and wind parameters.  This result in turn provides strong guidance for further physical modeling of the winds and X-ray generation.  

This conclusion may appear surprising, yet it was foreshadowed by the strong correlation between the bolometric and X-ray luminosities of the OB stars that has been known for many years (Long \& White 1980, Sana et~al.\ 2006).  Moreover, several of the objects with peculiar X-ray spectra, such as those in Figure~6, deviated from that correlation (e.g., Cassinelli et~al.\ 1994, Cohen et~al.\ 1997).  Of course, further physical modeling that uniquely reproduces the morphological correlations is required in order to understand them.

\acknowledgments
This research made use of the \textit{Chandra} Transmission Grating
Catalogue and archive (Huenemoerder et al.\ 2009).  We thank Mr.\ Arik Mitschang for assistance with the line measurements and plots at the Center for Astrophysics.  We also thank Alceste Bonanos, Lida Oskinova, Myron Smith, Rick White, and an anonymous referee for useful comments.  This research was supported by grants GO5-6006 to the authors from the \textit{Chandra} X-Ray Observatory Center, in the context of \textit{Chandra} Cycle~6 Proposal No.\ 6200204 (PI WLW).  WLW also acknowledges partial support from SAO grants AR8-9003A and GO8-9021B.  \textit{Chandra} is operated by the Smithsonian Astrophysical Observatory under NASA contract NAS8-03060, which also provided publication support for this paper.

\clearpage

\begin{deluxetable}{llcccccccll}
\rotate
\tabletypesize{\tiny}
\tablewidth{0pt}
\tablecaption{Stellar and Observational Parameters}
\tablehead{
\colhead{Star} &\colhead{Spectral Type}
&\colhead{$T_{eff}$\tablenotemark{a}}
&\colhead{$vsini$\tablenotemark{b}} &\colhead{$E(B-V)$}
&\colhead{Normalization} 
&\colhead{Ne\,\textsc{x}/} &\colhead{$T_{Ne}$} &\colhead{Exp.\ Time}
&\colhead{PI} &\colhead{Refs.}\\
& &{[kK]} &{[km~s$^{-1}$]} & &{Factors\tablenotemark{c}}
&\multicolumn{1}{l}{Ne\,\textsc{ix}} &{[MK]} &{[days]}}
\startdata
\sidehead{Figures~1 and 3:}
HD~93250 &O3.5~V((f+)) &43.8 &100 &0.47 &1.08$-$4/1.92$-$4/1.56$-$4 &0.72 &4.73 &2.25 &Waldron &1\\
9~Sgr &O4~V((f)) &42.9 &150 &0.35 &1.56$-$4/6.69$-$4/8.68$-$4 &0.27 &3.63 &1.71 &Waldron &2\\
HD~206267 &O6.5~V((f))~+~O9.5:~V &37.9 &... &0.53
&1.56$-$4/3.74$-$4/7.97$-$4 &0.36 &3.94 &0.86 &Schulz &3\\
15~Mon &O7~V((f))~+~O9.5~Vn &36.9 &100 &0.07 &5.45$-$5/6.58$-$4/1.74$-$3 &0.08 &2.81 &1.16 &Waldron  &4\\ 
$\iota$~Ori &O9~III~+~? &31.8 &125 &0.06 &3.41$-$4/3.42$-$3/7.26$-$3
&0.18 &3.33 &0.58 &Canizares &5\\ 
$\sigma$~Ori~AB &O9.5~V~+~B0.5~V &31.9 &110 &0.06
&1.26$-$4/2.09$-$3/4.03$-$3 &0.12 &3.08 &1.06 &Skinner &6, 7\\
$\beta$~Cru &B0.5~III &30.0 &$\sim$40 &0.04 &7.62$-$5/2.10$-$3/1.15$-$2 &0.08 &2.81 &0.86 &Cohen &8\\
\sidehead{Figures~2 and 4:}
HD~150136 &O3.5~If*~+~O6 V &41.3 &160 &0.48 &5.47$-$4/9.61$-$4/9.47$-$4 &0.42 &4.08 &1.04 &Skinner &9, 10\\
$\zeta$~Pup &O4~I(n)f &40.4 &230 &0.06 &1.85$-$3/5.92$-$3/4.49$-$3 &0.37 &3.97 &0.79 &Cassinelli &11, 12, 13, 14\\
$\xi$~Per &O7.5~III(n)((f)) &35.0 &215 &0.33 &1.99$-$4/1.01$-$3/1.34$-$3
&0.26 &3.60 &1.84 &Massa &\nodata\\
$\tau$~CMa &O9~II &31.8 &110 &0.16 &6.09$-$5/3.53$-$4/1.44$-$3 &0.18
&3.33 &1.01 &Canizares &\nodata\\
$\delta$~Ori &O9.5~II~+~B0.5~III &30.8 &145 &0.09
&4.98$-$4/3.78$-$3/9.00$-$3 &0.21 &3.43 &0.57 &Cassinelli &15, 16\\
$\zeta$~Ori &O9.7~Ib &30.5 &135 &0.08 &6.82$-$4/5.52$-$3/1.02$-$2 &0.22
&3.46 &0.69 &Waldron &17, 18, 19, 20\\
$\epsilon$~Ori &B0~Ia &30.0 &91 &0.08 &4.15$-$4/3.28$-$3/8.25$-$3 &0.22
&3.46 &1.06 &Waldron &21\\
\sidehead{Figure 6:}
$\theta^1$~Ori~C &O7~Vp &36.9 &130 &0.32 &3.68$-$3 &0.87 &5.04 &4.59
&Canizares, Gagn\'e &22, 23\\
$\tau$~Sco &B0.2~V &30.0 &$\leq$5 &0.04 &1.23$-$2 &0.42 &4.08 &0.83
&Cohen &24, 25\\
$\zeta$~Oph &O9.5~Vnn &31.9 &385 &0.32 &4.00$-$3 &0.43 &4.11 &0.97
&Waldron &26, 27\\
$\gamma$~Cas &B0.5~IVe~var &30.0 &380 &0.08 &5.72$-$3 &1.64 &6.23 &0.60 &Smith &28\\
\enddata
\tablenotetext{a}{Martins et al. 2005; 30~kK adopted for B0--B0.5.}
\tablenotetext{b}{Howarth \& Prinja 1989, Howarth et al. 1997, Harmanec
2002 ($\gamma$~Cas), SIMBAD ($\beta$~Cru).}
\tablenotetext{c}{Peak emission fluxes in 
photons~s$^{-1}$~cm$^{-2}$~{\AA}$^{-1}$ by
which the spectrograms have been divided to normalize them in the plots.
The negative powers of ten follow the minus signs.  The first value
pertains to wavelengths below 11~\AA, the second for 11--18~\AA, and
the third above 18~\AA\ in Figures~3 and 4; whereas, only the largest (or
sole) value applies to the entire wavelength ranges in Figures~1, 2, (and 6).}
\tablerefs{
References to Table~1:  1, Evans et~al.\ 2004; 2, Rauw et~al.\ 2002; 
3, Burkholder et~al.\ 1997; 4, Gies et~al.\ 1997; 5, Gies et~al.\ 1996; 6, Skinner et~al.\ 2008; 7, Sanz-Forcada et~al.\ 2004; 8, Cohen et~al.\ 2008; 9, Skinner et~al.\ 2005; 10, Niemela \& Gamen 2005; 11, Leutenegger et~al.\ 2007; 12, Kramer et~al.\ 2003; 13, Cassinelli et~al.\ 2001; 14, Kahn et~al.\ 2001; 15, Harvin et~al.\ 2002; 16, Miller et~al.\ 2002; 17, Raassen et~al.\ 2008; 18, Pollock 2007; 19, Cohen et~al.\ 2006; 20, Waldron \& 
Cassinelli 2001; 21, Waldron 2005a; 22, Gagn\'e et~al.\ 2005; 23, Schulz
et~al.\ 2003; 24, Cohen et~al.\ 2003; 25, Mewe et~al.\ 2003; 26, Li et~al.\ 2008; 27, Waldron 2005b; 28, Smith et~al.\ 2004.
}
\tablecomments{
The references in the table are to studies of individual stars, in inverse 
chronological order for a given star.  Analyses of many or most stars in
the table have also been presented by Waldron \& Cassinelli 2007, Zhekov \& Palla 2007, Wojdowski \& Schulz 2005, and other references cited in the text.
}
\end{deluxetable}

\newpage

\begin{deluxetable}{lcrrrclcrclcl}
\rotate
\tabletypesize{\footnotesize}
\tablenum{2}
\tablewidth{0pt}
\tablecaption{Rest Wavelengths of X-Ray Lines [\AA]}
\tablehead{
\multicolumn{5}{c}{He-like fir Lines} &&\multicolumn{3}{c}{H-like Lines}
&&\multicolumn{3}{c}{Strongest Fe\,\textsc{xvii} Lines}\\
\cline{1-5} \cline{7-9} \cline{11-13}
\noalign{\vspace{3pt}}
\colhead{Ion} &\colhead{IP [eV]} &\multicolumn{3}{c}{Wavelength}
&&\colhead{Ion} &\colhead{IP [eV]} &\colhead{Wavelength}
&&\colhead{Ion} &\colhead{IP [eV]} &\colhead{Wavelength}\\
\cline{3-5}
&&\colhead{r-line} &\colhead{i-line} &\colhead{f-line}
}
\startdata
O\,\textsc{vii} &\phn739 &21.602 &21.801, 21.804 &22.098
&&N\,\textsc{vii} &\phn667 &24.779, 24.785  &&Fe\,\textsc{xvii} &1266 &17.096\\
Ne\,\textsc{ix} &1196 &13.447 &13.550, 13.553 &13.699
&&O\,\textsc{viii} &\phn871 &18.967, 18.973 &&Fe\,\textsc{xvii} &1266 &17.051\\
Mg\,\textsc{xi} &1762 &9.169 &9.228, \phn9.231 &9.314
&&Ne\,\textsc{x} &1362 &12.132, 12.138      &&Fe\,\textsc{xvii} &1266 &16.780\\
Si\,\textsc{xiii} &2438 &6.648 &6.685, \phn6.688 &6.740
&&Mg\,\textsc{xii} &1963 &8.419, \phn8.425 &&Fe\,\textsc{xvii} &1266 &15.996 (blend)\\
S\,\textsc{xv} &3224 &5.039 &5.063, \phn5.066 &5.102
&&Si\,\textsc{xiv} &2673 &6.180, \phn6.186  &&Fe\,\textsc{xviii} &1358 &16.004 (blend)\\
&&&&
&&S\,\textsc{xvi} &3494 &4.727, \phn4.733   &&O\,\textsc{viii} &\phn871 &16.006 (blend)\\
&&&&&&&&
                      &&Fe\,\textsc{xvii} &1266 &15.261\\
&&&&&&&&
                      &&Fe\,\textsc{xvii} &1266 &15.014\\
&&&&&&&& 
                      &&Fe\,\textsc{xvii} &1266 &12.266\\
&&&&&&&&
                      &&Fe\,\textsc{xvii} &1266 &12.124 (Ne\,\textsc{x} blend)\\
\enddata
\end{deluxetable}

\clearpage

\begin{figure}
\epsscale{0.9}
\plotone{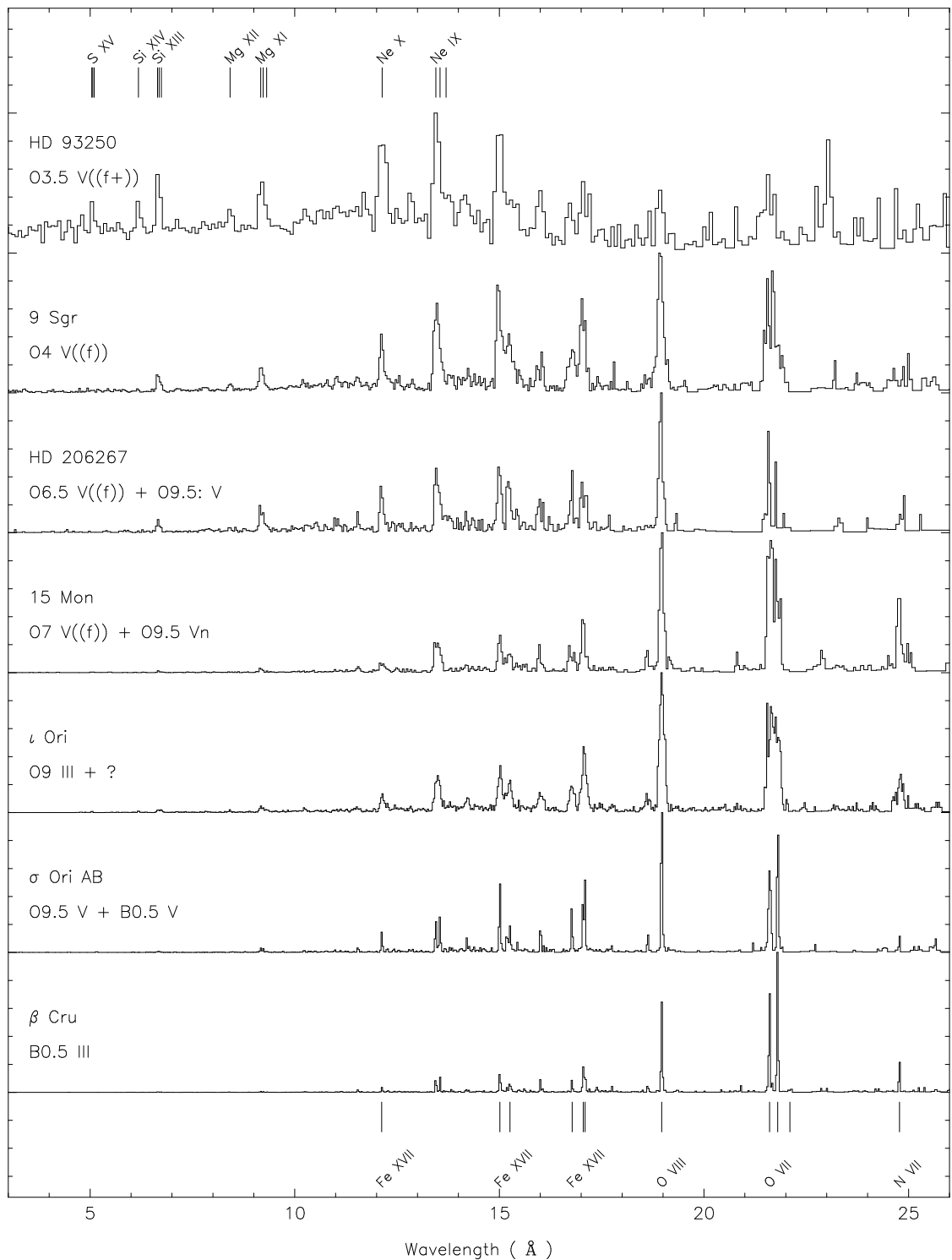}
\caption{\label{fig:fig1}
X-ray spectral-type sequence of normal dwarfs and giants.  The ordinate
units are normalized to 1.0 at the peak emission feature in each
spectrum, and the corresponding factors are given in Table~1.  Thus,
this scale may not be intercompared among different spectrograms.  The 
wavelengths of the identified spectral lines are given in Table~2.}
\end{figure}

\begin{figure}
\epsscale{0.9}
\plotone{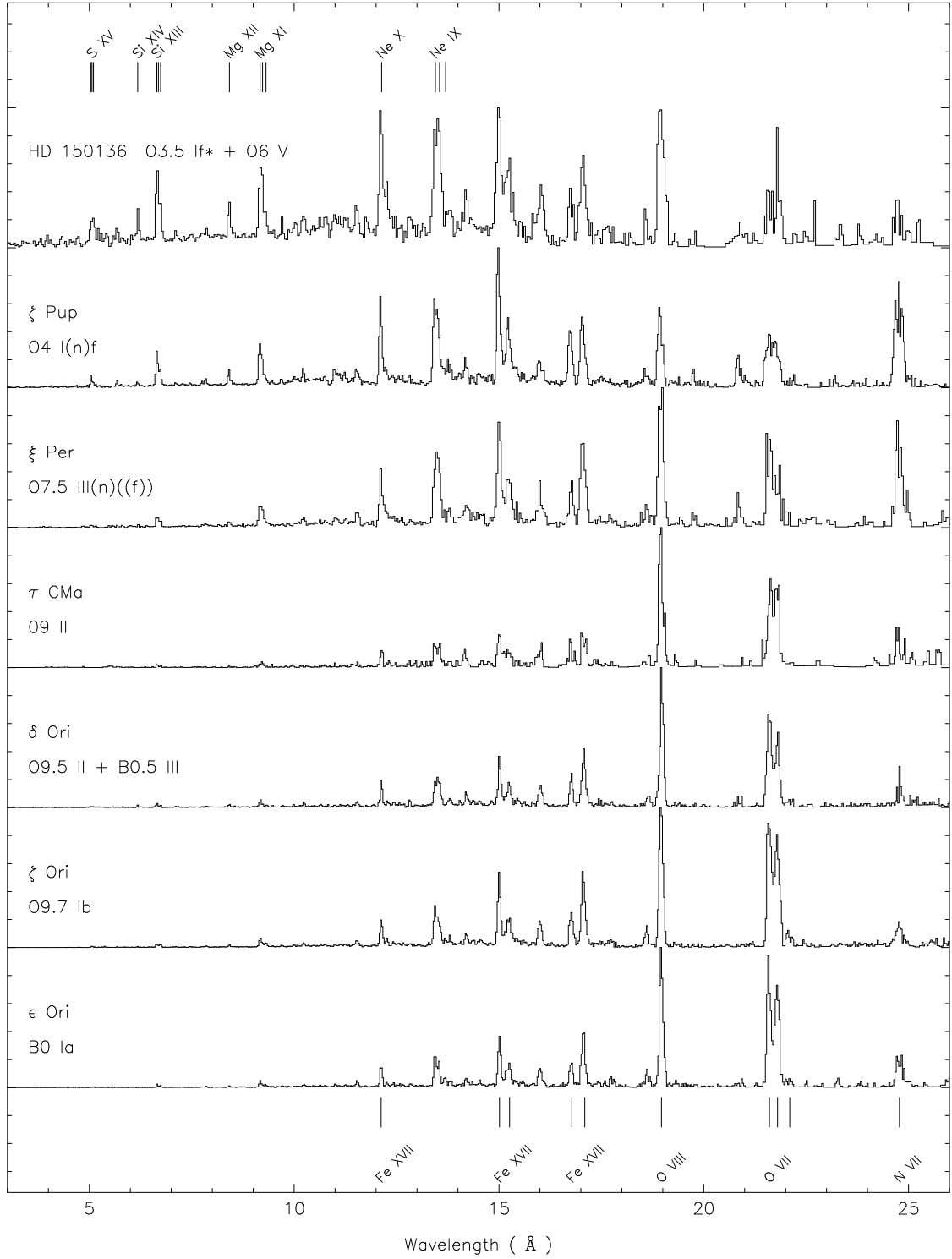}
\caption{\label{fig:fig2}
X-ray spectral-type sequence of normal supergiants and giants.  The ordinate units and the wavelengths of the identified spectral lines are as in Fig.~1.}
\end{figure}

\begin{figure}
\epsscale{0.9}
\plotone{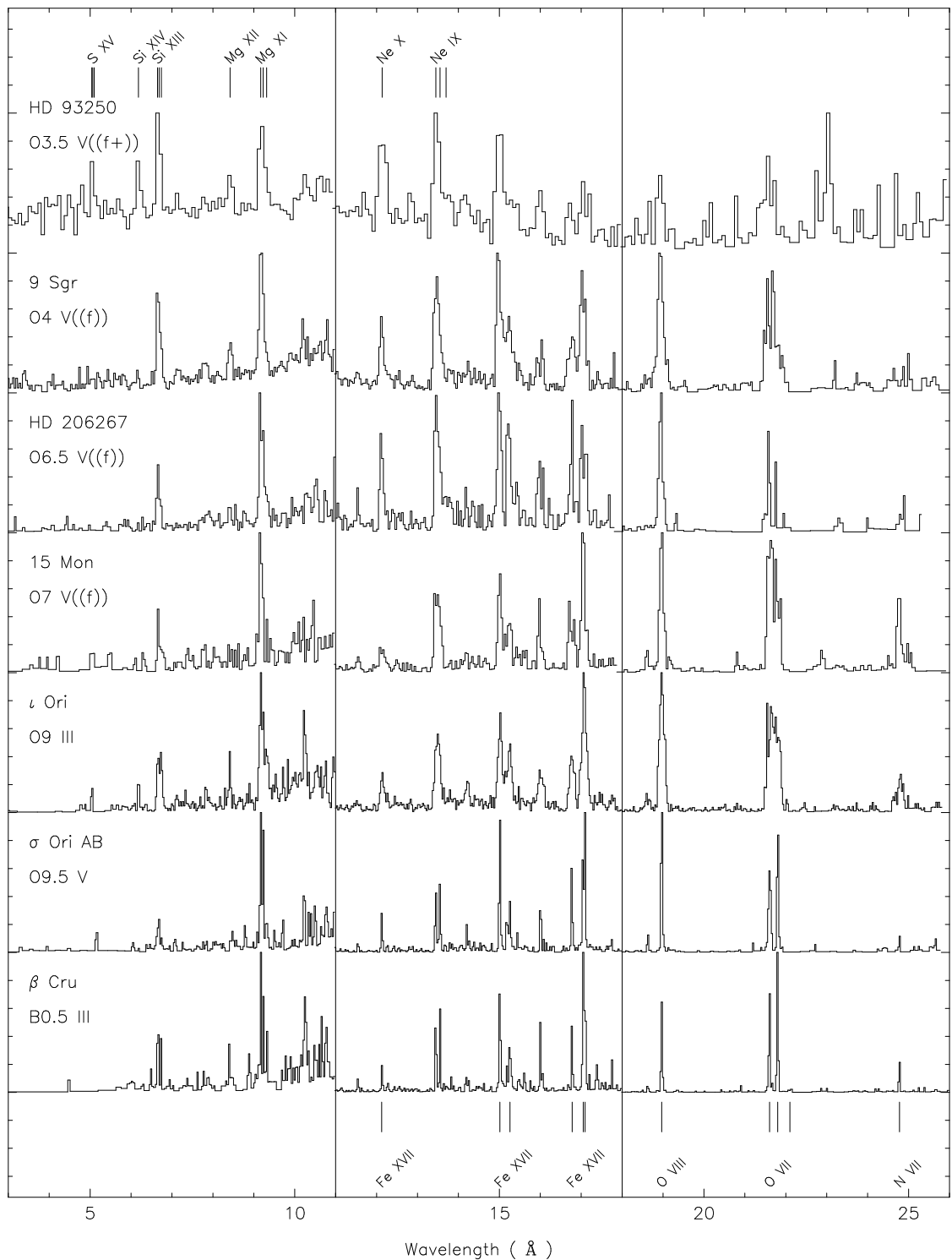}
\caption{\label{fig:fig3}
The same X-ray spectral-type sequence of normal dwarfs and giants as in
Fig.~1, except that the ordinate units are normalized to 1.0 by different
factors in each of the three panels, at the peak emission feature in each. 
The corresponding factors are given in Table~1.  The wavelengths of the 
identified spectral lines are given in Table~2.}
\end{figure}

\begin{figure}
\epsscale{0.9}
\plotone{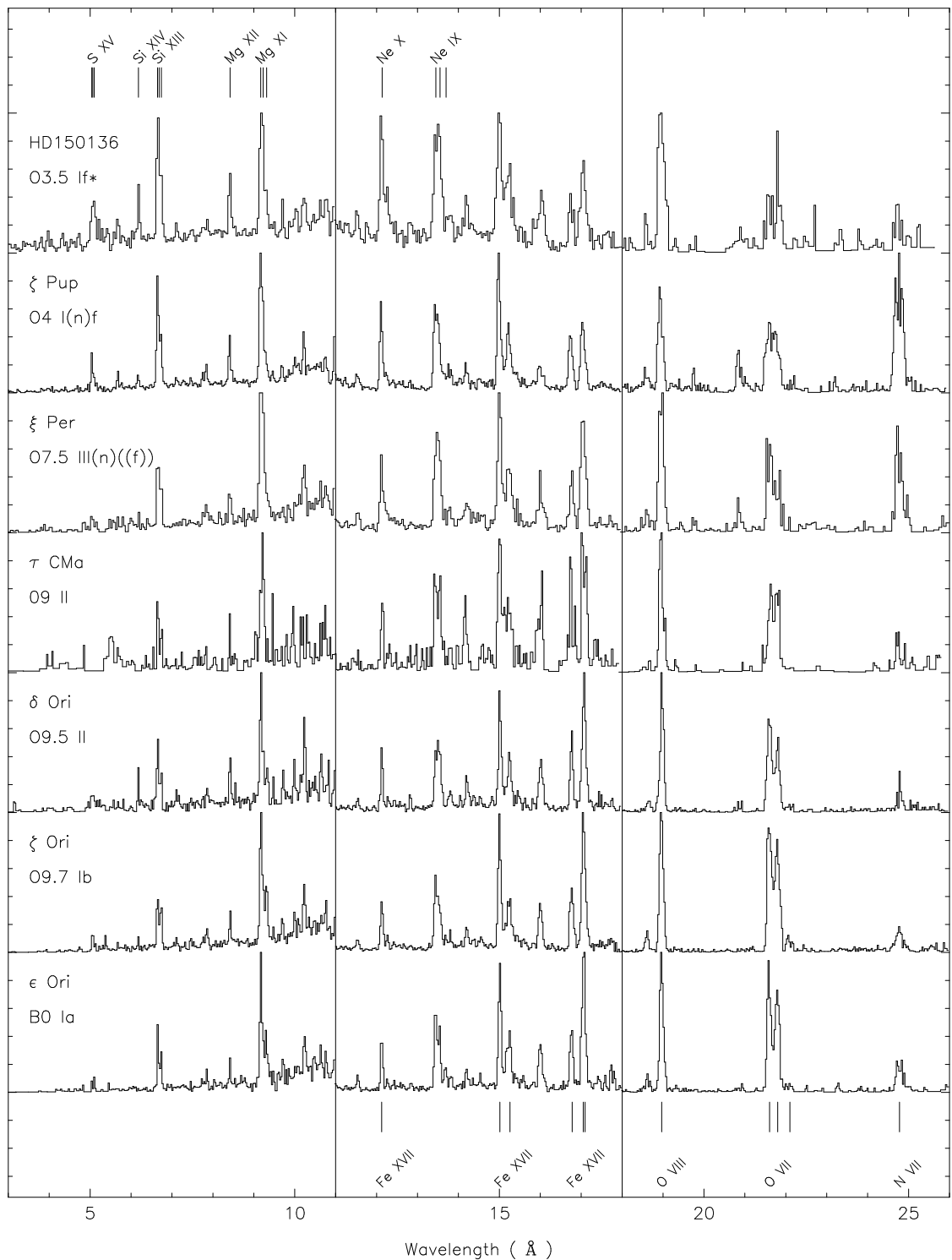}
\caption{\label{fig:fig4}
The same X-ray spectral-type sequence of normal supergiants and giants as
in Fig.~2, except with different normalization factors in each panel, as
in Fig.~3.  The wavelengths of the identified spectral lines are given in 
Table~2.}
\end{figure}

\begin{figure}
\epsscale{0.8}
\plotone{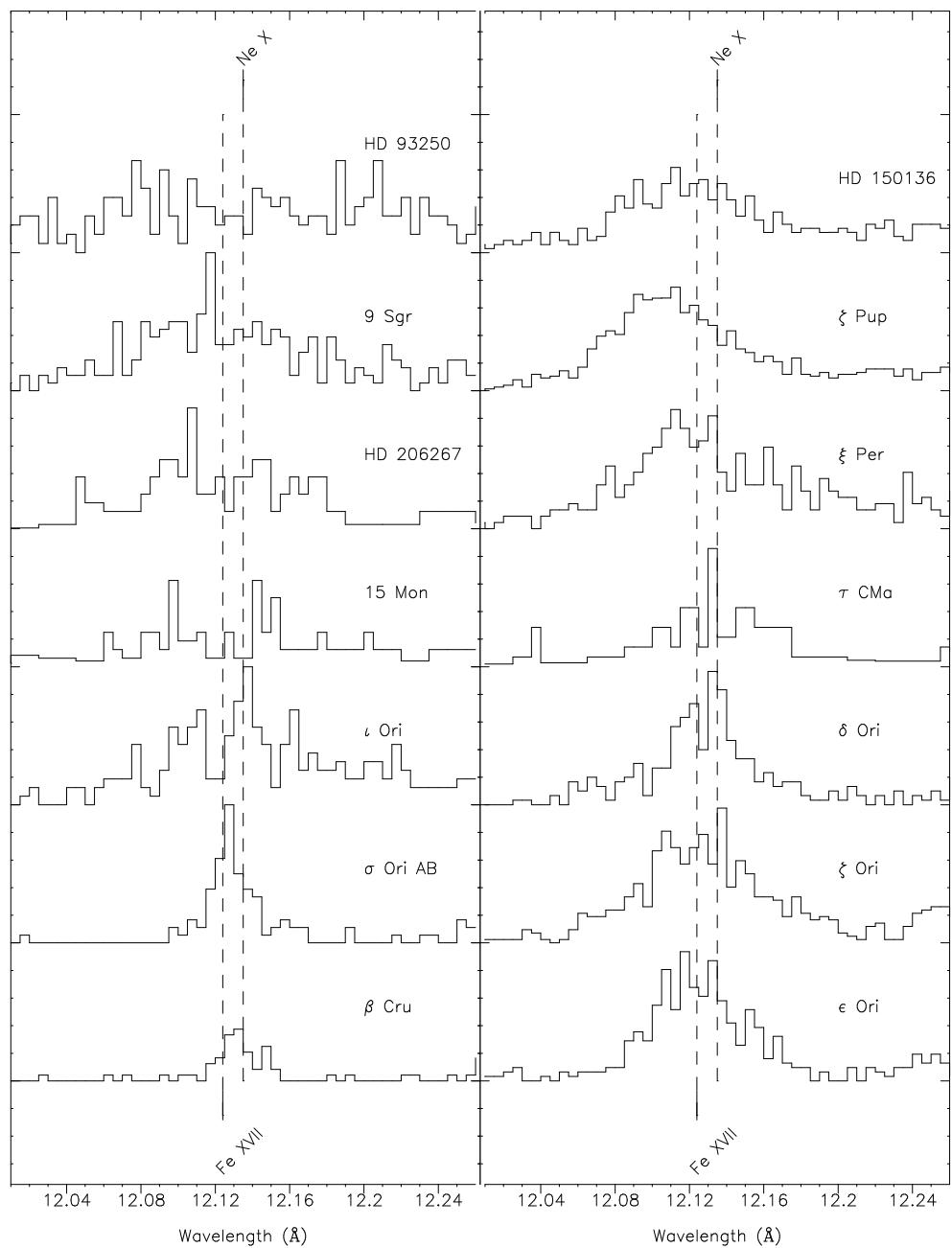}
\caption{\label{fig:fig5}
An expanded wavelength scale for the Ne\,\textsc{x} feature, blended with an Fe\,\textsc{xvii} line, in all the spectra of Figures~1--4.  While the two features cannot be resolved, the increasing relative contribution of the Fe line at a slightly shorter wavelength (Table~2) in the later spectral types 
can be discerned.  The Ne feature is very broad in the two earliest dwarfs, 
and it is blueshifted in the two earliest supergiants.}
\end{figure}

\begin{figure}
\epsscale{0.8}
\plotone{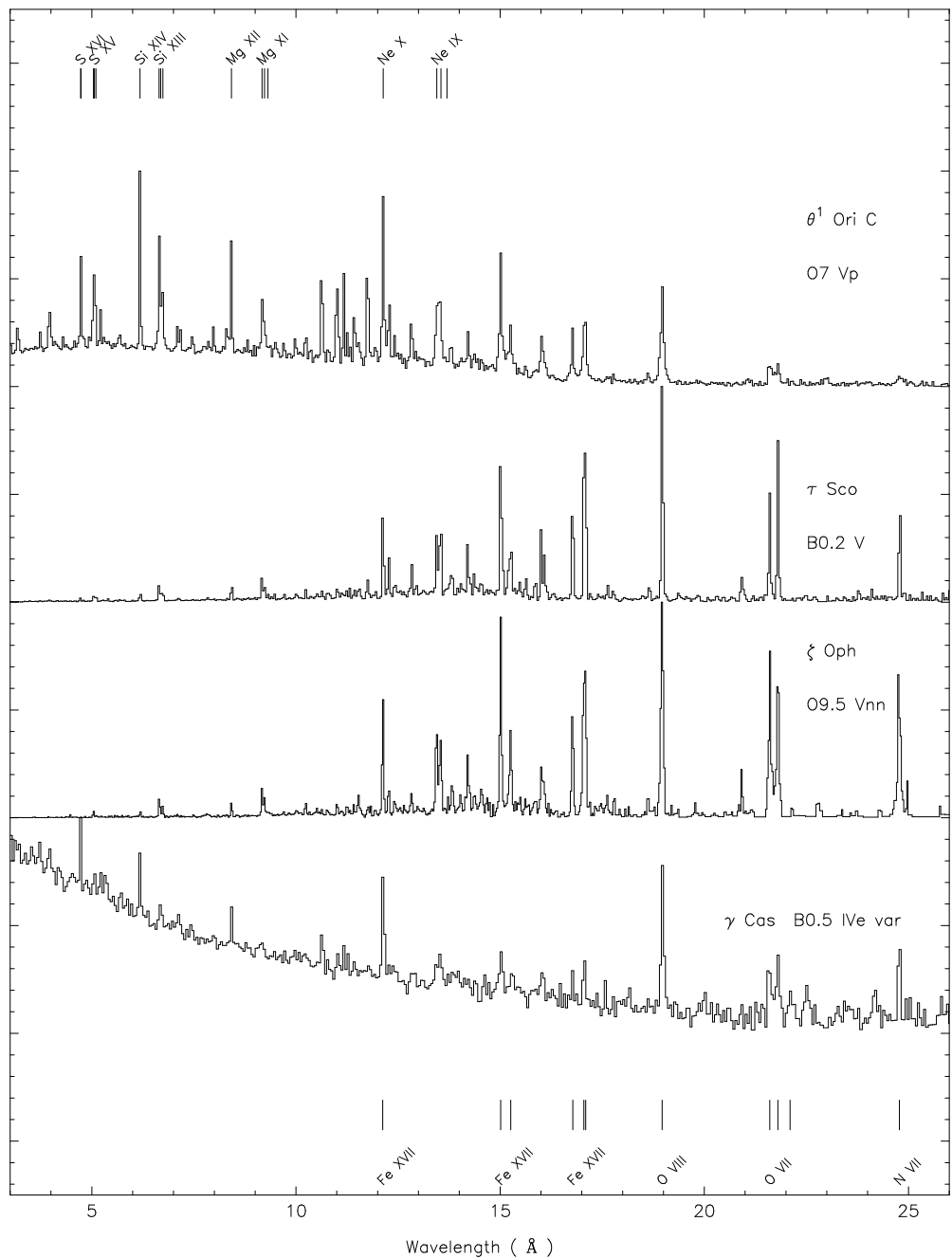}
\caption{\label{fig:fig6}
X-ray spectrograms of the magnetic stars $\theta^1$~Ori~C (Donati et~al.\ 2002, Gagn\'e et~al.\ 2005, Wojdowski \& Schulz 2005, Smith \& Fullerton 2005, Stahl et~al.\ 2008) and $\tau$~Sco (Cohen et~al.\ 2003, Wojdowski \& Schulz 2005, Donati et~al.\ 2006); and the rapid rotators $\zeta$~Oph (Waldron 2005b, Li et~al.\ 2008) and $\gamma$~Cas (Smith et~al.\ 2004).  The H/He-like line ratios are much larger than in normal stars of the same spectral types. The ordinate units and the wavelengths of the identified spectral lines are as in Fig.~1.}
\end{figure}

\begin{figure}
\epsscale{1.0}
\plotone{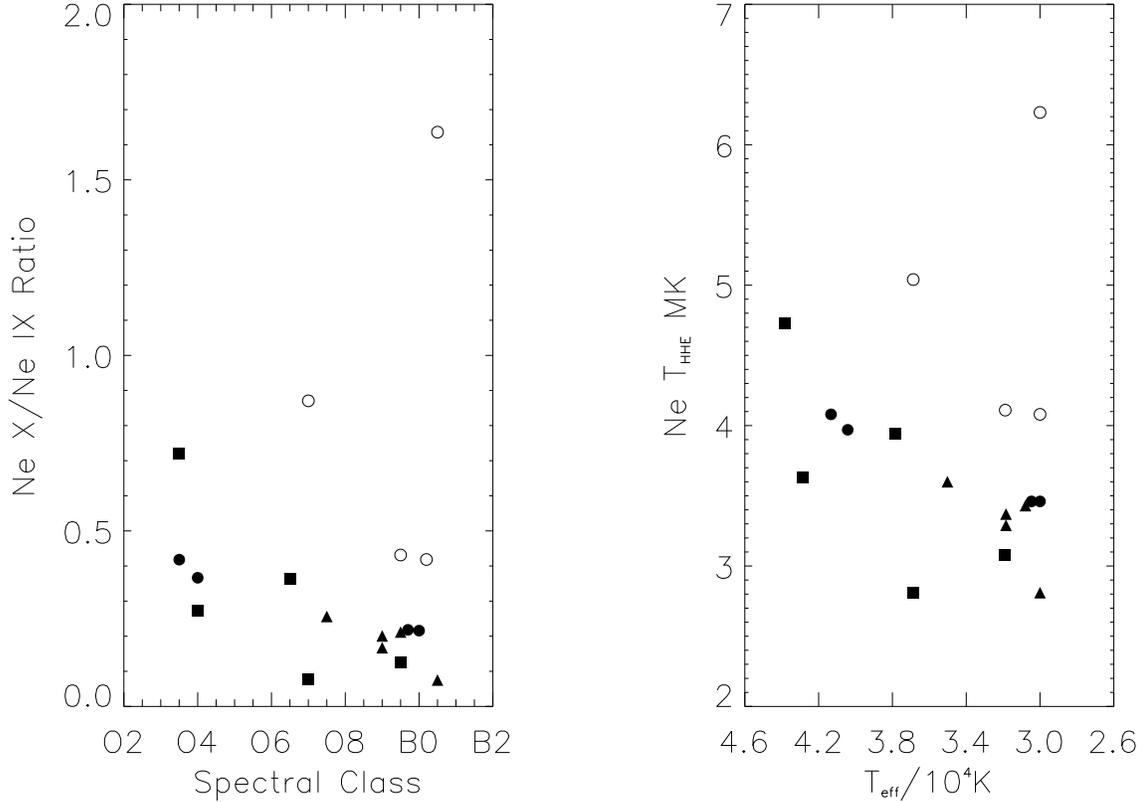}
\caption{\label{fig:fig7}
{\it Left}: Plot of the Ne\,\textsc{x}/Ne\,\textsc{ix} ratios as a 
function of spectral type.  The symbols are as follows: 
\textit{filled squares}, luminosity class V; \textit{filled triangles}, 
classes III and II; \textit{filled circles}, class I; and \textit{open 
circles}, peculiar objects.  The symbols for $\iota$~Ori and $\tau$~CMa have been offset by $\pm 0.018$ vertically, to avoid complete superposition. 
{\it Right}: Analogous plot of the X-ray temperatures derived from the
Ne ratios vs. the effective temperatures corresponding to the spectral
types (Table~1).  $\iota$~Ori and $\tau$~CMa have been offset by $\pm 0.025$ vertically.  In each plot, the normal stars show a declining X-ray ionization trend with advancing type or effective temperature, while the peculiar objects have higher ionization than normal stars of the same types or temperatures.}
\end{figure}

\end{document}